\documentclass[twocolumns,bibyear]{aa}
\usepackage{graphicx}
\usepackage{epsfig}
\usepackage{amssymb,amsmath}
\usepackage{txfonts}

\usepackage{natbib}

\usepackage{pslatex}
\usepackage{caption}
\usepackage{longtable}
\usepackage{epsfig}
\usepackage{pslatex}
\usepackage{color}

\newcommand\data{x^{\rm d}}
\newcommand{\neff}{N_{\rm eff}}
\newcommand{\eV}{\,{\rm eV}}

\begin{document}

\title{Cosmology with  gamma-ray bursts: II Cosmography challenges and cosmological scenarios for the accelerated
Universe }
\titlerunning{Cosmography challenges and dark energy}
\authorrunning{Demianski et al.}
\author{Marek Demianski$^{1,2}$,  Ester Piedipalumbo$^{3,4}$, Disha Sawant$^{5,6}$, Lorenzo Amati$^{7}$}
\institute{$^{1}$Institute for Theoretical Physics, University of Warsaw,  Pasteura 5, 02-093 Warsaw, Poland. \\
$^{2}$Department of Astronomy, Williams College, Williamstown, MA 01267, USA.\\
$^{3}$Dipartimento di Fisica, Universit\`{a} degli Studi di Napoli Federico II, Compl. Univ. Monte S. Angelo, 80126 Naples, Italy\\
\email{ester@na.infn.it}. \\
$^{4}$ I.N.F.N., Sez. di Napoli, Compl. Univ. Monte S. Angelo, Edificio 6, via Cinthia, 80126 - Napoli, Italy.\\
$^{5}$Department of Physics and Earth Sciences, University of Ferrara, Block C, Via Saragat 1, Ferrara, Italy 41122\\
$^{6}$Department of Physics, University of Nice Sophia Antipolis, Parc Valrose 06034, Nice Cedex2, France\\
$^{7}$INAF-IASF, Sezione di Bologna, via Gobetti 101, 40129 Bologna, Italy. }
\date{Accepted xxx, Received yyy, in original form zzz}
\abstract
 {Explaining the accelerated expansion of the Universe is one of the fundamental challenges in physics today.
 Cosmography provides information about
the evolution of the universe derived from measured distances, assuming only that the space time
geometry is described by the Friedman-Lemaitre-Robertson-Walker metric, and adopting an approach that
effectively uses only  Taylor expansions of basic observables.}
 {We perform a high-redshift analysis  to constrain  the cosmographic expansion up to the fifth order.
 It is based on the Union2 type Ia supernovae data set, the  gamma-ray burst Hubble diagram, a data set
 of 28 independent measurements of the Hubble parameter, baryon acoustic oscillations measurements from galaxy clustering
and the Lyman-$\alpha$ forest  in the SDSS-III Baryon Oscillation Spectroscopic Survey (BOSS), and some
Gaussian  priors on $h$ and $\Omega_M$.}
{We performed a statistical analysis and explored the probability
distributions of the cosmographic parameters. By building up their regions of confidence,  we maximized our
likelihood function using the Markov chain Monte Carlo method.}
{Our high-redshift analysis  confirms that the expansion of the Universe currently accelerates; the estimation
of the jerk  parameter indicates a possible deviation from the standard $\Lambda$CDM cosmological model. Moreover,
we  investigate implications of our results for the reconstruction of the dark energy equation of state (EOS) by
comparing  the standard  technique of cosmography with an alternative approach based on generalized
Pad\'e approximations of the same observables. Because these expansions converge better,  is possible to improve
the constraints on the cosmographic parameters and also on the dark matter EOS.}
{The estimation of the jerk
and the DE parameters indicates at $ 1 \sigma $  a possible deviation from the $\Lambda$CDM cosmological
model.}

 \date{}
\maketitle
\section{Introduction}
In the past dozen years a huge and diverse set
of observational data revealed that the Universe is now expanding at an accelerated rate,
see for instance
\cite{Riess07}, \cite{SNLS}, \cite{Riess11}, \cite{Union2.1}, \cite{PlanckXXVI}, and
\cite{PlanckXIII}. It is usually
assumed that this accelerated expansion is caused by the so-called dark energy,
a cosmic medium  with unusual properties. The pressure of dark energy $p_{de}$ is
negative, and it is related to the positive energy density of dark energy
$\epsilon_{de}$ by the equation of state (EOS), $p_{de}=w\epsilon_{de}$, where
the proportionality coefficient is $ < -1/3$. According to current estimates,
about 75\% of the matter energy in the Universe is in the form of dark energy, so
that today the dark energy is the dominant component in the Universe. The nature
of dark energy is not known. Models of dark energy proposed so far include at
least a non-zero cosmological constant (in this case $w=-1$), a potential energy
of some scalar field, effects connected with inhomogeneous distribution of matter
and averaging procedures, and extended theories of gravity  (an accelerated
expansion can be obtained by generalizing the Einstein theory of gravity  to some
theory derived from a modified action with respect to the Hilbert-Einstein
action: the simplest extension of General Relativity is achieved assuming that
the gravitational Lagrangian is  an arbitrary continuous function $f(R)$ of the
Ricci scalar $R$ . In this case, in general, $w\not= -1$ and it is not constant
and depends on the redshift $z$. Extracting the information on  the EOS of dark
energy from observational data is then at the same time a fundamental problem and
a challenging task. To probe the dynamical evolution of dark energy in these
circumstances, we can parameterize $w$ empirically, usually using two or more
free parameters. Of all the parametrization forms of the dark energy  EOS, the
Chevallier-Polarski-Linder (CPL) model \cite{cpl1}, \cite{cpl2}
is probably the most
widely used, since it presents  a smooth and bounded behavior for high redshifts
and a manageable two-dimensional parameter space and also provides a simple
and effective instrument of computations. However, it would result a in
physically incomplete parametrization of dark energy if we were to take into
account the inhomogeneities of the late-time Universe. Linear parametrizations of
the dark energy EOS (the CPL  EOS is linear in the scale factor $a$) are not
compatible with the theory of scalar perturbations in the late Universe.
Therefore these EOS are not the fundamental and can only be used to approximate
the real EOS \cite{akarsu15}. In our approach, this model is only used to
investigate whether the EOS is constant, independently of any assumption on the
nature of the DE: according to this point of view, even the small number of
parameters of the CPL  model is not as important as this
independence (in some scalar field models of dark energy, the so-called
quintessence, first introduced in \cite{Peebles88a}, \cite{Peebles88b}, the scalar field
has one free parameter less than CPL). Moreover, it is worth noting that even
neglecting the inhomogeneities, several dark energy models considered so far
agree reasonably well with the observational data, so that, unless higher
precision probes of the expansion rate and the growth of structure are developed,
these different approaches cannot be distinguished. This degeneration suggests a
kinematical approach to the problem of cosmic acceleration, relying on
quantities that are not model dependent. The cosmographic approach is related to
the derivatives of the scale factor and enables fitting the data on the distance
- redshift relation without any a priori assumption on the underlying
cosmological model. It is based on the sole assumption that the Universe is
spatially homogeneous and isotropic, and that it can be described by  the
Friedman-Lemaitre-Robertson-Walker (FLRW) metric. In our high-redshift  investigation,
extended behind the supernova type Ia (SNIa) Hubble diagram,  we require at
least a fifth-order Taylor expansion of the scale factor to obtain a reliable
approximation of the distance - redshift relation. As a consequence, it is in
principle possible to estimate up to five cosmographic parameters, $(h, q_0, j_0,
s_0, l_0)$, although the available high-redshift data sets are still too small
and do not allow us to obtain a precise and realistic determination of all of
them, see \cite{salzcosmo}. When these quantities have been determined, we can
use them to set constraints on the dark energy models.To constrain the
cosmographic parameters, we use the Union2 SNIa data set, the gamma-ray burst
(GRB) Hubble diagram, constructed  by calibrating the correlation between the
peak photon energy, $E_{\mathrm{p, i}}$, and the isotropic equivalent radiated
energy, $ E_{\mathrm{iso}}$ (see Paper I),  a sample of 28  measurements of the
Hubble parameter, compiled in \cite{farooqb},  Gaussian priors on the distance
from the baryon acoustic oscillations (BAO), and the Hubble constant $h$ (these
priors have been included to help break the degeneracies of the model
parameters). Our statistical analysis is based on
Monte Carlo Markov Chain (MCMC)  simulations to simultaneously compute the full probability density functions
(PDFs) of all the parameters of interest. The structure of the paper is as
follows. In Sect. 2 we describe the basic elements of the cosmographic approach
and explicitly derive series expansions of the scale factor and other relevant
parameters. In Sect. 3 we describe the observational data sets that are used in
our analysis. In Sect. 4 we describe some details of our statistical analysis and
present results on cosmographic parameters obtained from three sets of data. In
Sect. 5 we present constraints on dark energy models that can be derived from our
analysis. General discussion of our results and conclusions are presented in
Sect. 6.

\section{Cosmography approach }
Cosmic acceleration is one of the most remarkable
problem in physics and cosmology.
However, it is worth noting that all the evidence for this late-time accelerated dynamics
appears in the context of an assumed cosmological scenario and cosmological model. Recently,
the cosmographic approach to cosmology  gained increasing interest because the intention is
to collect as much information as possible directly from observations (mainly measured distances),
without addressing issues such as which type of dark energy and dark
matter are required to satisfy the Einstein equation, but just assuming  the minimal priors of isotropy
and homogeneity. This means that the space-time geometry is described by the
FLRW line element
\begin{equation}
ds^2=-c^2dt^2+a^2(t)\left[\frac{dr^2}{1-kr^2}+r^2d\Omega^2\right]\,,
\end{equation}
where $a(t)$ is the scale factor and $k= +1, 0, - 1$ is the curvature parameter.
With this metric, it is possible to express the luminosity
distance $d_L$ as a power series in the redshift parameter $z$, the
coefficients of the expansion being functions of the scale factor
$a(t)$ and its higher order derivatives. This
expansion leads to a distance\,-\,redshift relation that only
relies on the assumption of the FLRW metric and is therefore fully model independent since it does not depend on the
particular form of the solution of cosmic evolution equations. To this aim,
it is convenient to introduce the cosmographic functions \cite{Visser}
\begin{eqnarray}
\label{eq:cosmopar}
H(t) &\equiv& + \frac{1}{a}\frac{da}{dt}\, ,
\\
q(t) &\equiv& - \frac{1}{a}\frac{d^{2}a}{dt^{2}}\frac{1}{H^{2}}\,
,
\\
j(t) &\equiv& + \frac{1}{a}\frac{d^{3}a}{dt^{3}}\frac{1}{H^{3}}\,
,
\\
s(t) &\equiv& + \frac{1}{a}\frac{d^{4}a}{dt^{4}}\frac{1}{H^{4}}\,
,
\\
l(t) &\equiv& + \frac{1}{a}\frac{d^{5}a}{dt^{5}}\frac{1}{H^{5}}\,.
\end{eqnarray}
The cosmographic parameters, which are commonly indicated as the Hubble,
deceleration, jerk, snap, and
lerk parameters, correspond to the functions evaluated at the present time $t_0$. Note that the
use of the jerk parameter to distinguish between different models was also
proposed in \cite{SF} in the context of the  statefinder
parametrization. Furthermore, it is possible
to relate the derivative of the Hubble parameter to the other
cosmographic parameters\,
\begin{eqnarray}
\dot{H} &=& -H^2 (1 + q) \ , \label{eq: hdot}
\\
\ddot{H} &=& H^3 (j + 3q + 2) \ , \label{eq: h2dot}
\\
d^3H/dt^3 &=& H^4 \left ( s - 4j - 3q (q + 4) - 6 \right ) \ ,
\label{eq: h3dot}
\\
d^4H/dt^4 &=& H^5 \left ( l - 5s + 10 (q + 2) j + 30 (q + 2) q +
24 \right )\, \label{eq: h4dot}
\end{eqnarray}
where a dot denotes derivative with respect to the cosmic time
$t$. With these definitions the series expansion to the fifth order in
time of the scale factor is
\begin{eqnarray}\label{eq:a_series}
\frac{a(t)}{a(t_{0})} &=& 1 + H_{0} (t-t_{0}) -\frac{q_{0}}{2}
H_{0}^{2} (t-t_{0})^{2} +\frac{j_{0}}{3!} H_{0}^{3} (t-t_{0})^{3}
\\ \nonumber &+&\frac{s_{0}}{4!} H_{0}^{4} (t-t_{0})^{4}+ \frac{l_{0}}{5!}
H_{0}^{5} (t-t_{0})^{5} +\emph{O}[(t-t_{0})^{6}]\,.
\end{eqnarray}
From Eq.(\ref{eq:a_series}), and  recalling that
the distance traveled by a photon that is emitted at
time $t_{*}$ and absorbed at the current epoch $t_{0}$ is
\begin{equation}
D = c \int dt = c (t_{0} - t_{*})\,,
\end{equation}
we can construct the series for the luminosity or angular-diameter distance, whose expansions is

\begin{eqnarray}\label{serielum1}
d_{L}(z) = \frac{c z}{H_{0}} \left( \mathcal{D}_{L}^{0} +
\mathcal{D}_{L}^{1} \ z + \mathcal{D}_{L}^{2} \ z^{2} +
\mathcal{D}_{L}^{3} \ z^{3} + \mathcal{D}_{L}^{4} \ z^{4} +
\emph{O}(z^{5}) \right)\,,
\end{eqnarray}
with {\setlength\arraycolsep{0.2pt}
\begin{eqnarray}\label{serieslum2}
\mathcal{D}_{L}^{0} &=& 1\,, \\
\mathcal{D}_{L}^{1} &=& - \frac{1}{2} \left(-1 + q_{0}\right)\,, \\
\mathcal{D}_{L}^{2} &=& - \frac{1}{6} \left(1 - q_{0} - 3q_{0}^{2} + j_{0}\right)\,, \\
\mathcal{D}_{L}^{3} &=& \frac{1}{24} \left(2 - 2 q_{0} - 15
q_{0}^{2} - 15 q_{0}^{3} + 5 j_{0} + 10 q_{0} j_{0} + s_{0} \right)\,,\\
\mathcal{D}_{L}^{4} &=& \frac{1}{120} \left( -6 + 6 q_{0} + 81
q_{0}^{2} + 165 q_{0}^{3}  -105 q_{0}^{4} - 110 q_{0} j_{0}  +\right.\nonumber \\ \nonumber - && \left.105
q_{0}^{2} j_{0} - 15 q_{0} s_{0} - 27 j_{0} + 10 j^{2} - 11 s_{0} - l_{0}\right)\,,
\end{eqnarray}}
and
\begin{eqnarray}
d_{A}(z) = \frac{c z}{H_{0}} \left ( \mathcal{D}_{A}^{0} +
\mathcal{D}_{A}^{1}  z + \mathcal{D}_{A}^{2} \ z^{2} +
\mathcal{D}_{A}^{3}  z^{3} + \mathcal{D}_{A}^{4}  z^{4} +
\emph{O}(z^{5}) \right )\,,
\end{eqnarray}
with{\setlength\arraycolsep{0.2pt}
\begin{eqnarray}
\mathcal{D}_{A}^{0} &=& 1 \,,\\
\mathcal{D}_{A}^{1} &=& - \frac{1}{2} \left(3 + q_{0}\right)\,, \\
\mathcal{D}_{A}^{2} &=& \frac{1}{6} \left(11 + 7 q_{0} + 3q_{0}^{2} - j_{0} \right)\,, \\
\mathcal{D}_{A}^{3} &=& - \frac{1}{24} \left(50 + 46 q_{0} + 39
q_{0}^{2} + 15 q_{0}^{3} - 13 j_{0} - 10 q_{0} j_{0} - s_{0} + \right. \\  \nonumber && \left.
- \frac{2 k c^{2} (5 + 3 q_{0})}{H_{0}^{2} a_{0}^{2}}\right)\,, \\
\mathcal{D}_{A}^{4} &=& \frac{1}{120} \left( 274 + 326 q_{0} + 411
q_{0}^{2} + 315 q_{0}^{3} + 105 q_{0}^{4}  - \right. \\ && \left.- 210 q_{0} j_{0} - 105
q_{0}^{2} j_{0}-15 q_{0} s_{0} +
 137 j_{0} + 10 j_0^{2} - 21 s_{0} - l_{0} \right)\,.\nonumber
\end{eqnarray}}
It is worth noting that since the cosmography is based on series expansions, the fundamental difficulties
of applying this approach to fit the luminosity distance data using high-redshift distance indicators are
connected with the
convergence and truncation of the series. Recently, the possibility of attenuating the convergence problem
has been analyzed by defining a new redshift variable, see \cite{vitagliano}, the so-called y-redshift,
\begin{equation}
z \rightarrow y = \frac{z}{1+z}~.
\end{equation}
For a series expansion in the classical z-redshift the
convergence radius is equal to $1$, which is a drawback
when the application of cosmography is to be extended to
redshifts $z>1$.
The y-redshift might help to solve this problem because the z-interval
$[0,\infty]$ corresponds to the y-interval $[0,1]$, so that we are
mainly inside the convergence interval of the series, even for Cosmic Microwave Background
data ($z = 1089 \rightarrow y = 0.999$). In principle, we
might therefore extend the series up to the redshift of decoupling, and place CMB-related constraints
within the cosmographic approach.
However, even using the series expansions in y-redshift, the problem of the series truncation remains
(see also \cite{zhang16}).
The higher the order of the cosmographic expansion, the more accurate the approximation. However, as we add
cosmographic parameters, the volume of the parameter space increases and the constraining strength could be
weakened by degeneracy effects among different parameters. Therefore, the order of truncation
depends on a compromise of different requirements. To fix a reliable expansion
order of the cosmographic series, we first performed a qualitative analysis by fixing a fiducial model,
given by the recently released Planck data \cite{PlanckXIII}, that is, a flat quintessence  model,
characterized by $\Omega_m=0.315 \pm 0.017$ and $w= -1.13^{+ 0.17}_{-0.10}$. The dimensionless Hubble function,
$E(z)$ associated with this model is
\begin{equation}\label{Hquint}
E(z)=\sqrt{\Omega_m \left(1+z\right)^3+ \Omega_{\Lambda}^{3\left(1+w \right)}}.
\end{equation}
This model was used to construct a mock high-redshift
Hubble diagram data set: we realized 500 simulations by
randomly extracting the fiducial model parameters in their error range, and we also used the distribution
of the most updated GRB Hubble diagram. For any redshift value we evaluated the mean and the dispersion of
the distribution of the distance modulus, which characterize a normal probability function, from which
the Hubble diagram data points are picked up.
Finally, the exact values of the cosmographic parameters, derived from Eq. (\ref{Hquint}), were compared
with the corresponding values of cosmographic series up to the fourth order, fitted on our mock data  set.
A significant degeneracy was detected
in that even well-constrained cosmological parameters can correspond to larger uncertainties of the
cosmographic parameters, which increase for higher order terms. This degeneracy can be only partially
attributed to the accuracy of the cosmographic reconstruction: only $q_0$ and $j_0$ are well constrained.
In the analysis we considered a forth-order expansion and were able to successfully set bounds on these
parameters in a statistically  consistent way.

It is worth to stress that the  GRB Hubble diagram spans an optimal  redshift range for
the sensitivity of the observables quantities on the cosmological parameters,
with special attention on the cosmography and its implications on dark energy. We
show this in Fig. 1, following a simplified approach, in which we consider the
distance modulus $\mu(z)$ as observable: we fixed a flat $\Lambda$CDM
fiducial cosmological model by  constructing the corresponding
$\mu_{fid}(z, \rm{\theta})$,  and plot the percentage error on the distance modulus with respect to
different corresponding functions randomly generated within an evolving CPL EOS.
The higher sensitivity is only reached for $z\geqslant 3$, that is, a redshift
region unexplored by SNIa and BAO samples.

To provide reasonably narrow statistical constraints, we applied an MCMC method
that allowed us to obtain marginalized likelihoods on the series coefficients,
from which we infer tight constraints on these parameters. We have inserted
several tests in our code that give us control over several physical requirements
we expect from the theory. For instance, since we use data related to the Hubble
parameter $H(z)$, we are able to set restrictions on the Hubble parameter,
$H_0=H(0)$, and thus to obtain a considerable improvement in the quality of
constraints.

\begin{figure}
\includegraphics[width=7 cm]{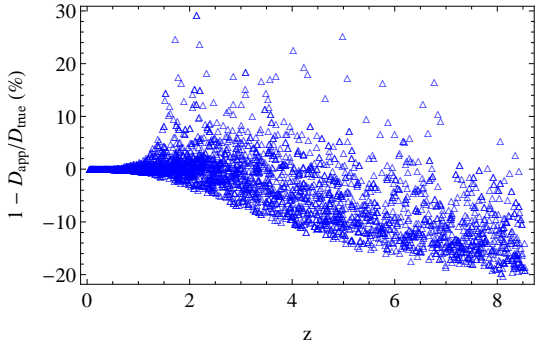}
        \caption{We plot the z dependence of the percentage error in the distance modulus between the fiducial $\Lambda$CDM model and different flat models, with an evolving EOS $w(z)=w_0+ w_1 \displaystyle \frac{z}{1+z}$:
        the GRB Hubble diagram  span a better suited redshift than the SNIa range ($z <  2 $).}
        \label{depar}
\end{figure}

\section{Observational data sets}
In our cosmographic approach we use  the currently available observational data sets on SNIa and GRB Hubble diagram,
and we set Gaussian priors on the distance data from the BAO and the Hubble constant $h$. These priors were
included to help break the degeneracies of the parameters of the cosmographic series expansion in Eqs. (\ref{serielum1}).
\subsection{Supernovae and GRB Hubble diagram}

\subsubsection{Supernovae Ia}
 In the past decade the confidence in type Ia supernovae as standard candles has steadily
grown.   SNIa observations gave the first strong indication of the recently accelerating
expansion of the Universe. Since 1995, two teams of astronomers, the High-Z
Supernova Search Team and the Supernova Cosmology Project, have been discovering type Ia supernovae at high
redshifts. First results of the teams were published by  \cite{Riess} and
\cite{per+al99}.
Here we consider the recently updated Supernovae Cosmology Project Union 2.1 compilation \cite{Union2.1},
which is an update of the original Union compilation and contains $580$ SNIa, spanning the redshift
range ($0.015 \le z \le 1.4$).
We compare the theoretically predicted distance modulus $\mu(z)$
with the observed one through a Bayesian approach, based on the definition
of the cosmographic distance modulus,

\begin{equation}
\mu(z_{j}) = 5 \log_{10} ( D_{L}(z_{j}, \{\theta_{i}\}) )+\mu_0\,,
\end{equation}
where $D_{L}(z_{j}, \{\theta_{i}\})$ is the Hubble free luminosity
distance, expressed as a series depending on the cosmographic parameters, $\theta_{i}=(q_{0},
j_{0}, s_{0}, l_{0})$.  The parameter $\mu_{0}$ encodes the Hubble
constant and the absolute magnitude $M$, and has to be
marginalized over. Given the heterogeneous origin of the Union data
set, we worked with an alternative version of  the $\chi^2$:
\begin{equation}
\label{eq: sn_chi_mod}
\tilde{\chi}^{2}_{\mathrm{SN}}(\{\theta_{i}\}) = c_{1} -
\frac{c^{2}_{2}}{c_{3}}\,,
\end{equation}
where
\begin{equation}
c_{1} = \sum^{{\cal{N}}_{SNIa}}_{j = 1} \frac{(\mu(z_{j}; \mu_{0}=0,
\{\theta_{i})\} -
\mu_{obs}(z_{j}))^{2}}{\sigma^{2}_{\mathrm{\mu},j}}\, ,
\end{equation}
\begin{equation}
c_{2} = \sum^{{\cal{N}}_{SNIa}}_{j = 1} \frac{(\mu(z_{j}; \mu_{0}=0,
\{\theta_{i})\} -
\mu_{obs}(z_{j}))}{\sigma^{2}_{\mathrm{\mu},j}}\, ,
\end{equation}
\begin{equation}
c_{3} = \sum^{{\cal{N}}_{SNIa}}_{j = 1}
\frac{1}{\sigma^{2}_{\mathrm{\mu},j}}\,.
\end{equation}
It is worth noting that
\begin{equation}
\chi^{2}_{\mathrm{SN}}(\mu_{0}, \{\theta_{i}\}) = c_{1} - 2 c_{2}
\mu_{0} + c_{3} \mu^{2}_{0} \,,
\end{equation}
which clearly becomes minimum for $\mu_{0} = c_{2}/c_{3}$, so
 that $\tilde{\chi}^{2}_{\mathrm{SN}} \equiv
\chi^{2}_{\mathrm{SN}}(\mu_{0} = c_{2}/c_{3}, \{\theta_{i}\})$.
\subsubsection{Gamma-ray burst Hubble diagram}
Gamma-ray bursts are visible up to high redshifts thanks to the enormous energy
that they release, and thus may be good candidates  for our high-redshift
cosmological investigation. However, GRBs may be everything but standard candles
since their peak luminosity spans a wide range, even if  there have been many
efforts to make them distance indicators using some empirical correlations of
distance-dependent quantities and rest-frame observables  \cite{Amati08}. These
empirical relations allow us to deduce the GRB rest-frame luminosity or energy
from an observer-frame measured quantity, so that the distance modulus can be
obtained with an error that depends essentially on the intrinsic scatter of the
adopted correlation. We performed our cosmographic analysis using a GRB Hubble diagram data
set, built by calibrating the $E_{\rm p,i}$ -- $E_{\rm iso}$ relation. We recall
that  $E_{\rm iso}$ cannot be measured directly, but can be obtained through the
knowledge of the bolometric fluence, denoted by $ S_{bolo}$. This is more
correctly $E_{\rm iso}=4\pi d^2_{L}(z)S_{\rm bolo}(1+z)^{-1}\,$. Therefore $
E_{\rm iso}$ depends  on the GRB observable, $S_{\rm bolo}$, but also on the
cosmological parameters. At first glance, it seems that the calibration of these
empirical laws depends on the assumed cosmological model. To use GRBs as tools
for cosmology, this circularity problem has to be overcome, see for
instance,  \cite{Li08}, \cite{gao}, \cite{Liang08}, \cite{Samushia}, \cite{Liu14},
\cite{Wang15}, and \cite{wangrev}. In  Paper
I we have applied a local regression technique to estimate in a model-independent
way the distance modulus from the Union SNIa sample.

\begin{figure}
\includegraphics[width=7 cm]{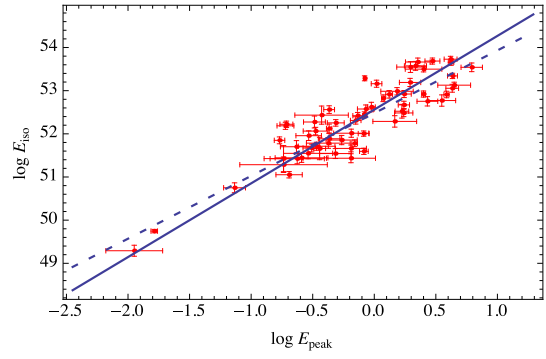}
\caption{Best-fit curves for the  $E_{\rm p,i}$ -- $E_{\rm iso}$ correlation relation superimposed on
the data. The solid and dashed lines refer to the results
obtained with the maximum likelihood (Reichart likelihood) and weighted $\chi^2$ estimator,
respectively.}
\label{eg-episocorreich}
\end{figure}
\begin{figure}
\includegraphics[width=7 cm]{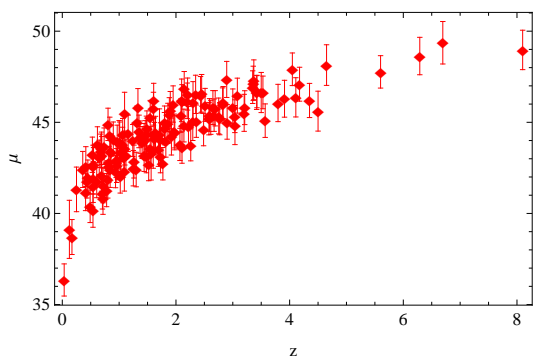}
\caption{Distance modulus $\mu(z)$ for the calibrated  GRB Hubble diagram obtained by fitting the $E_{\rm p,i}$ -- $E_{\rm iso}$ relation.}
\label{hdreich06}
\end{figure}

 When the  correlation is fit  (see Fig. \ref{eg-episocorreich}) and its parameters are estimated, it is possible
to compute the luminosity distance of a certain GRB at  redshift $z$  and, therefore, estimate the distance modulus
for each $i$\,-\,th GRB in our sample at redshift
$z_i$, and to build the Hubble diagram plotted in Fig. \ref{hdreich06}.

\subsection{Baryon acoustic oscillations  data}
Baryon acoustic oscillations data are promising standard rulers to investigate
different cosmological scenarios and models. They  are related to density
fluctuations induced by acoustic waves that are created by primordial
perturbations:  the peaks of the acoustic waves gave rise to denser regions in
the distribution of baryons, which, at recombination,  imprint the correlation
between matter densities at the scale of the sound horizon.  Measurements of
CMB radiation provide the absolute
physical scale for these baryonic peaks, but the observed position of the peaks
of the two-point correlation function of the matter distribution, compared with
such absolute values, enables  measuring cosmological distance scales. To use
BAOs as a cosmological tool, we follow \cite{P10} and define\,

\begin{equation}
d_z = \frac{r_s(z_d)}{d_V(z)}\,,
\label{eq: defdz}
\end{equation}
where $z_d$ is the drag redshift computed with the approximated formula in \cite{EH98}, $r_s(z)$ is the
comoving sound horizon\,,
\begin{equation}
r_s(z) = \frac{c}{\sqrt{3}} \int_{0}^{(1 + z)^{-1}}{\frac{da}{a^2 H(a) \sqrt{1 + (3/4) \Omega_b/\Omega_{\gamma}}}} \ ,
\label{defsoundhor}
\end{equation}
and $d_V(z)$ the volume distance, that is,\,
 \begin{equation}
 d_V(z) = \left[\left(1+z\right) d_A(z)^2\frac{c z}{H(z)}\right]^{\frac{1}{3}}.\label{volumed}
\end{equation}
Here  $ d_A(z)$ is the angular diameter distance.
Moreover, BAO measurements in spectroscopic surveys allow directly estimating the expansion rate H(z),
converted into the quantity $D_{H}(z)=\displaystyle\frac{c}{H(z)}$, and constraints (from transverse
clustering)  on the comoving angular diameter distance $D_{M}(z)$, which in a flat FLRW metric is
$D_{M}(z)\propto c \int^{z}_{0 }\frac{d\zeta}{H(\zeta)}$. To perform our  analysis using BAO data,
all these distances were properly developed in terms of the corresponding cosmographic series.
The BAO data used in our analysis are summarized in Table \ref{tab:data} and are taken from  \cite{Aubourg14}.
\begin{table*}
  \centering
  \setlength{\tabcolsep}{1em}
  \begin{tabular}{|cccc|}
  \hline\hline
    Redshift &  $D_V/r_d$         &   $D_M/r_d$ &   $D_H/r_d$  \\
    \hline

                        0.106   &  $3.047 \pm 0.137$ &       --    &  --                       \\

                         0.15     &  $4.480 \pm 0.168$ &        --         & --     \\

    0.32    &  $8.467 \pm 0.167$  &       --     & --   \\
 0.57    &    --                    &  $14.945 \pm 0.210$ & $20.75 \pm 0.73 $  \\

2.34 &    --               &  $37.675 \pm 2.171$ & $9.18 \pm 0.28 $ \\
 2.36 & --             &  $36.288 \pm 1.344$ & $9.00 \pm 0.30 $  \\
    2.34    &    --               &  $36.489 \pm 1.152$ & $9.145 \pm 0.204 $ \\
\hline
  \end{tabular}

  \caption{
    BAO data used in our analysis.
  }
  \label{tab:data}
\end{table*}
Here, the BAO scale $r_d$  is the radius of the sound horizon at the
decoupling era, which can be approximated as
\begin{equation}
  r_d \approx {56.067\, \exp\left[-49.7 (\omega_\nu+0.002)^2\right]
    \over
    \omega_{cb}^{0.2436} \,
    \omega_b^{0.128876} \, [1+(\neff-3.046)/30.60] } ~{\rm Mpc}~,
  \label{eqn:rdneff}
\end{equation}
for a standard radiation background with
$\neff=3.046$, $\sum m_\nu < 0.6\eV$ $\omega_{\nu} = 0.0107{\sum m_\nu}/1.0 eV$, see  \cite{Aubourg14}.
Using the values of $\omega_b$ and $\omega_{cb}$ derived by Planck, we find that
$r_d = 147.49 \pm 0.59 Mpc$. It is worth noting that $\omega_{cb}$ indicates
the $\omega$ density of the baryons + CDM.
\subsection{H(z) measurements}
\begin{figure}
\includegraphics[width=7 cm]{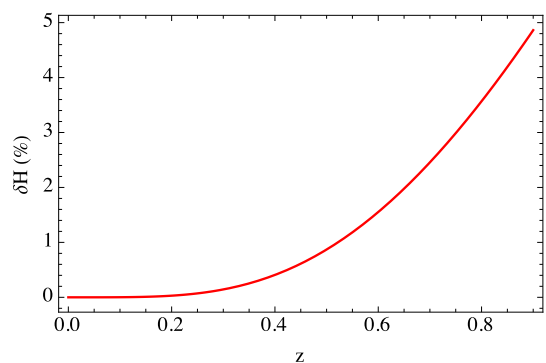}
\caption{Relative error between the exact Hubble parameter $H(z)$ in the standard flat $\Lambda$CDM model  ($\Omega_m= 0.3$, $h=0.7$)  and the corresponding cosmographic approximation: we need to limit our dataset
at $z\leq 0.9$ when we wish to maintain an accuracy of a few $\text{percent}$ .}
\label{deltaH}
\end{figure}

 The measurements of Hubble parameters are a complementary probe to constrain the cosmological parameters and
 investigate the dark energy \cite{farooqa}, \cite{farooqb}. The Hubble parameter, defined as
 $H(z) = \displaystyle \frac{\dot a}{a}$,  where $a$ is the scale factor, depends on the differential
 age of the Universe as a function of redshift and can be measured  using the so-called  cosmic chronometers.
  $dz$ is obtained from spectroscopic surveys with high accuracy, and  the differential evolution of the age of
  the Universe $dt$ in the  redshift interval $dz$ can be measured provided that  optimal probes of the aging of
  the Universe, that is, the cosmic chronometers, are identified \cite{moresco}. The most reliable cosmic
  chronometers at present
are old early-type galaxies that evolve passively on a timescale much longer than their age difference, which
formed the vast majority of their stars rapidly and  early  and have not experienced  subsequent major episodes
of star formation.
Moreover, the Hubble parameter can also be obtained from the BAO measurements: by observing the typical acoustic
scale in the light-of-sight direction, it is possible to extract the expansion rate of the Universe at a certain
redshift \cite{busca}. We used a list of $28$ $H(z)$ measurements, compiled in \cite{farooqb} and shown in
Table (\ref{hz}).
\begin{table}
\begin{center}
 \centering
  \setlength{\tabcolsep}{1em}
\begin{tabular}{|c|c|c|}
\hline\hline
~~$z$ & ~~$H(z)$ &~~~~~~~ $\sigma_{H}$ \\
~~~~~    & (km s$^{-1}$ Mpc $^{-1}$) &~~~~~~~ (km s$^{-1}$ Mpc $^{-1}$) \\
\hline\hline
0.070&~~        69&~~~~~~~      19.6\\
0.100&~~        69&~~~~~~~      12\\
0.120&~~        68.6&~~~~~~~    26.2\\
0.170&~~        83&~~~~~~~      8\\
0.179&~~        75&~~~~~~~      4\\
0.199&~~        75&~~~~~~~      5\\
0.200&~~        72.9&~~~~~~~    29.6\\
0.270&~~        77&~~~~~~~      14\\
0.280&~~        88.8&~~~~~~~    36.6\\
0.350&~~        76.3&~~~~~~~    5.6\\
0.352&~~        83&~~~~~~~      14\\
0.400&~~        95&~~~~~~~      17\\
0.440&~~        82.6&~~~~~~~    7.8\\
0.480&~~        97&~~~~~~~      62\\
0.593&~~        104&~~~~~~~     13\\
0.600&~~        87.9&~~~~~~~    6.1\\
0.680&~~        92&~~~~~~~      8\\
0.730&~~        97.3&~~~~~~~    7.0\\
0.781&~~        105&~~~~~~~     12\\
0.875&~~        125&~~~~~~~     17\\
0.880&~~        90&~~~~~~~      40\\
0.900&~~        117&~~~~~~~     23\\
1.037&~~        154&~~~~~~~     20\\
1.300&~~        168&~~~~~~~     17\\
1.430&~~        177&~~~~~~~     18\\
1.530&~~        140&~~~~~~~     14\\
1.750&~~        202&~~~~~~~     40\\
2.300&~~        224&~~~~~~~     8\\

\hline\hline
\end{tabular}
\end{center}
\caption{Hubble parameter versus redshift data, as compiled in \cite{farooqb}
}\label{hz}
\end{table}
To also achieve a high accuracy approximation in terms of the proper cosmographic series for  the $H(z)$,
we decided to consider only data with $z< 0.9$. We found that $H(z)$ is much more sensitive to the order of
the approximation and to the values of the cosmographic parameters than any distance observables. The relative
error on $H(z)$, $\delta H$, remains on the order of few percents only in this redshift range, as we show in
Fig. \ref{deltaH}.

\section{Statistical analysis}
To constrain the cosmographic parameters, we performed a preliminary and standard fitting procedure
to maximize the likelihood function ${\cal{L}}({\bf p})$. This requires the knowledge of the precision matrix,
that is, the inverse of the covariance matrix of the measurements,

\begin{eqnarray}
\footnotesize
{\cal{L}}({\bf p}) & \propto & \frac {\exp{(-\chi^2_{SNIa/GRB}/2)}}{(2 \pi)^{\frac{{\cal{N}}_{SNIa/GRB}}{2}} |{\bf C}_{SNIa/GRB}|^{1/2}}  \frac{\exp{(-\chi^2_{BAO}/2})}{(2 \pi)^{{\cal{N}}_{BAO}/2} |{\bf C}_{BAO}|^{1/2}} \nonumber\\ ~ & \times & \frac{1}{\sqrt{2 \pi \sigma_{\omega_m}^2}}
 \exp{\left [ - \frac{1}{2} \left ( \frac{\omega_m - \omega^{obs}_{m}}{\sigma_{\omega_m}} \right )^2\right ]} ,\\ &&\times\frac{1}{\sqrt{2 \pi \sigma_h^2}} \exp{\left[ - \frac{1}{2} \left ( \frac{h - h_{obs}}{\sigma_h} \right )^2\right]} \frac{\exp{(-\chi^2_{H}/2})}{(2 \pi)^{{\cal{N}}_{H}/2} |{\bf C}_{H}|^{1/2}}\nonumber
 \\ & \times & \frac{1}{\sqrt{2 \pi \sigma_{{\cal{R}}}^2}} \exp{\left [ - \frac{1}{2} \left ( \frac{{\cal{R}} - {\cal{R}}_{obs}}{\sigma_{{\cal{R}}}} \right )^2 \right ]} \nonumber \,
\label{defchiall}
\end{eqnarray}

where

\begin{equation}
\chi^2(\bf p) = \sum_{i,j=1}^{N} \left( \data_i - x^{th}_i(p)\right)C^{-1}_{ij}  \left( \data_j - x^{th}_j(p)\right) \,.
\label{eq:chisq}
\end{equation}
Here $\bf p$ is the set of parameters, $N$ is the number of data points,  $\bf \data_i$ is the $i-th$ measurement;
$\bf x^{th}_i(p)$
indicate the theoretical predictions for these measurements and depend on the parameters $\bf p$;  $\bf C_{ij}$ is
the covariance matrix (specifically,
${\bf C}_{SNIa/GRB/H}$ indicates the SNIa/GRBs/H  covariance matrix);
$(h^{obs}, \sigma_h) = (0.742, 0.036)$\, \cite{shoes},
and $(\omega_{m}^{obs}, \sigma_{\omega_m}) = (0.1356, 0.0034)$\, \cite{PlanckXIII}.
It is worth noting that  the effect of our prior on h is not critical at all so that we are
certain that our results are not biased by this choice. There are two
opinions on $h$, one that claims it is centered on $h=0.68$, and the other on $h=0.74$\,
\cite{chen}, \cite{sievers}, \cite{hinshaw}, \cite{Aubourg14}, and \cite{PlanckXIII}.  In any event,
we recall that
the strongest dependence of the constraints on the assumed value of $H_0 $ is for the
H(z) data alone.
The term
$\displaystyle  \frac{1}{\sqrt{2 \pi \sigma_{{\cal{R}}}^2}} \exp{\left [ - \frac{1}{2} \left ( \frac{{\cal{R}} - {\cal{R}}_{obs}}{\sigma_{{\cal{R}}}} \right )^2 \right ]} $
in the likelihood (\ref{defchiall}) considers the shift parameter ${\cal{ R}}$:

\begin{equation}
{\cal{R}} = H_{0} \sqrt{\Omega_M} \int_{0}^{z_{\star}}{\frac{dz'}{H(z')}}\,,
\label{eq: defshiftpar}
\end{equation}
where $z_\star = 1090.10$ is the redshift of the surface of last scattering  \cite{B97}, \cite{EB99}\,. According to
the Planck data  $({\cal{R}}_{obs}, \sigma_{{\cal{R}}}) = ( 1.7407, 0.0094)$.

\begin{figure}
\includegraphics[width=9 cm, height=4cm]{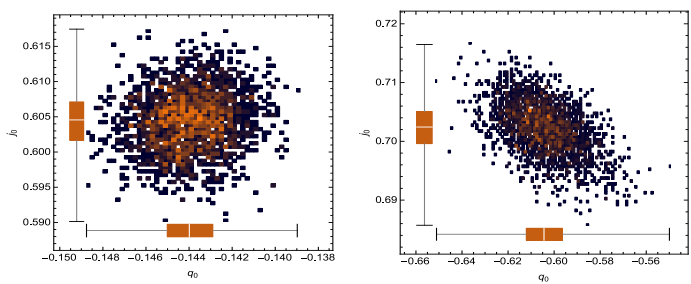}
\caption{ Confidence regions in the ($q_0$-$j_0$) plane, as provided by  Cosmography Ib and II. The inner brown region defines the 2\,$\sigma$ confidence level. The parameters  $q_0$ and $j_0$ are well constrained, the values $q_0>0$ are ruled out, the value $j_0=1$ (which is the  $\Lambda$CDM value) is statistically not favorable.  On the axes we also plot the box-and-whisker diagrams for the  respective parameters: the bottom and top of the diagrams are the 25th and 75th percentile (the lower and upper quartiles, respectively), and the band near the middle of the box is the median.} \label{conregq0j0}
\end{figure}

Finally, the term  $\displaystyle \frac{\exp{(-\chi^2_{H}/2})}{(2
\pi)^{{\cal{N}}_{H}/2} |{\bf C}_{H}|^{1/2}}  $ in Eq. (\ref{defchiall}) takes
into account some recent measurements of $H(z)$ from the differential age of
passively evolving elliptical galaxies. We used the data collected by (Stern et
al. 2010) giving the values of the Hubble parameter for ${\cal{N}}_H = 11$
different points over the redshift range $0.10 \le z \le 1.75$ with a diagonal
covariance matrix. We finally performed our cosmographic analysis by considering
a whole data set containing all the data sets, which we call the
cosmographic data set. To  sample the ${\cal{N}} $ dimensional space of
parameters, we used the  MCMC method and ran five parallel chains and used the
Gelman - Rubin diagnostic approach to test the convergence. As a test probe, it
uses the reduction factor $R$, which is the square root of the ratio of
the variance between-chain  and the variance within-chain.  A large $R$
indicates that the between-chain variance is substantially greater than the
within-chain variance, so that a longer simulation is needed. We require that
$R$  converges to 1 for each parameter. We set $R - 1$ of order $0.05$,
which is more restrictive than the often used and recommended value $R - 1 < 0.1$
for standard cosmological investigations. Moreover, to reduce the uncertainties
of the cosmographic  parameters, since methods like the MCMC are based on an
algorithm that moves randomly in the parameter space, we a priori impose
in the code some basic consistency controls requiring that all of the
(numerically) evaluated values of $H(z)$ and $d_{L}(z)$ be positive. As first
step,  we performed a sort of pre-statistical analysis to
select the starting points of the full analysis:  we ran our chains to
compute the likelihood in Eq. (\ref{defchiall}) considering only the SNIa.
Therefore we applied  the same  MCMC approach to evaluate the likelihood in Eq.
(\ref{defchiall}), combining the SNIa, H(z), and the BAO  data with the GRBs Hubble diagram,
as described above. We discarded the  first $30\%$  of the point iterations at
the beginning of any MCMC run, and thinned the chains that were run many times.
We finally   extracted the constraints on cosmographic parameters by coadding the
thinned chains.  The histograms of the parameters from the merged chains were
then used to infer median values and confidence ranges: the $15.87$th and
$84.13$th quantiles define the $68 \%$ confidence interval; the $2.28$th and
$97.72$th quantiles define the $95\%$ confidence interval;  and the $0.13$th and
$99.87$th quantiles define the $99\%$ confidence interval. In Tables \ref{tab1},
\ref{tabgrb}, and \ref{tabfull} we present the results of our analysis. Only the
deceleration parameter, $q_0$, and the jerk, $j_0$, are well constrained (see
also Fig \ref{conregq0j0}); the snap parameter, $s_0$, is weakly constrained, and
the lerk, $l_0$, is unconstrained, as has been found in the literature (see
\cite{MEcosmo}, \cite{escamillarivera}).

\begin{table}
\begin{center}
 \centering
  \setlength{\tabcolsep}{0.1 em}
\begin{tabular}{|c|c|c|c|c|c|}
  \hline
\hline
  Parameter&$h$&$q_0$&$j_0$&$s_0$\\
  \hline
  \hline
  Best fit & $0.74$&$-0.48$ & $0.68$ & $-0.51$ \\
  Mean & $0.74$&$ -0.48$&$0.65$&$-6.8$ \\
  2 $\sigma$ & $ (0.68, 0.72)$&$ (-0.5, -0.38)$&$(0.29, 0.98)$&$(-1.33, -0.53)$\\
  \hline
  \hline
\end{tabular}
\end{center}
\caption{Constraints on the cosmographic parameters from
combining the SNIa  Hubble diagram with the BAO and $H(z)$ data sets (Cosmography Ia).}
\label{tab1}
\end{table}

\begin{table}
\begin{center}
 \centering
  \setlength{\tabcolsep}{0.1 em}
\begin{tabular}{|c|c|c|c|c|c|}
  \hline
\hline
  Parameter&$h$&$q_0$&$j_0$&$s_0$\\
  \hline
  \hline
  Best fit & $0.67$&$-0.14$ & $0.6$ & $-5.55$ \\
  Mean & $0.67$&$ -0.14$&$0.6$&$-5.55$ \\
  2 $\sigma$ & $ (0.66, 0.73)$&$ (-0.15, -0.14)$&$(0.58,0.62)$&$(-5.7,6.1)$\\
  \hline
  \hline
\end{tabular}
\end{center}
\caption{Constraints on the cosmographic parameters  from
combining the GRBs Hubble diagram with the BAO and $H(z)$ data sets (Cosmography Ib).}
\label{tabgrb}
\end{table}

\begin{table}
\begin{center}
 \centering
  \setlength{\tabcolsep}{0.1 em}
\begin{tabular}{|c|c|c|c|c|c|}
 \hline
\hline
  Parameter &$h$&$q_0$&$j_0$&$s_0$\\
  \hline
  \hline
  Best fit & $0.72$&$-0.6$&$0.7$ & $-0.36$ \\
  Mean & $0.72$&$-0.6$&$0.7$&$-0.37$ \\
  2 $\sigma$ & $(0.67, 0.73)$&$ (-0.62, -0.55)$&$(0.69, 0.73)$&$(-0.4, 5)$\\
  \hline
  \hline
\end{tabular}
\end{center}
\caption{Constraints on the cosmographic parameters from
combining the SNIa and GRBs Hubble diagrams with the BAO  data set ( Cosmography II).}
\label{tabfull}
\end{table}

Our statistical MCMC analysis shows that the deceleration parameter
$q_0$ is clearly negative in all the cases. The marginal likelihood distributions for the current values
of the  deceleration parameter $q_0$ and  the jerk $j_0$  indicate an only negligible probability for $q_0>0$,
and  that $j_0$ is significantly different from its $\Lambda$CDM value $j_0=1$.
In Fig. \ref{conregq0j0} we show the confidence regions for $q_0$, and $j_0$: the left and the right
panels show the results obtained by using only the GRBs data set (Cosmography Ib) or the  whole sample
(Cosmography II);  the GRB data mainly constrain the jerk parameter.
\section{Connection with the  dark energy EOS}
As already mentioned, within the FLRW paradigm, all possibilities of interpreting
the dark energy  can be characterized, as far as the background dynamics is
concerned, by an effective  dark energy EOS $w(z)$.  Extracting information on
the EOS of dark energy from observational data is therefore   both an
issue of crucial importance and a challenging
task. To probe the dynamical evolution of dark energy, we can parameterize $w(z)$
empirically by assuming that it evolves smoothly with redshift and can be
approximated by some analytical expression, containing two or more free
parameters. Since any analytical form of w(z) is not based on a grounded theory,
in principle an extreme  flexibility is required, which means a large numbers of
parameters. However, at present,  the precision in the observational data is not
good enough to provide constraints for more than a few parameters (two or three
at most). Often, to reduce the huge arbitrariness, the space of allowed w(z)
models is restricted to consider only $w(z)\geq -1$. If  $w(z)$ is an
effective EOS parameterizing a modified gravity theory, for instance, a
scalar-tensor or an f(R) model, then this constraint might be too restrictive and
partially arbitrary.  Whereas we account the cosmography results as a sort of
constraint on the EOS space parameters. It is well known that the connection
between the cosmographic and the dark energy parametrization is based on the
series expansion of the Hubble function $H(z)$. For a spatially flat cosmological
model we have

\begin{eqnarray}
  H(z) &=& H_0 \sqrt{(1-\Omega_m) g(z)+\Omega_m (z+1)^3}\,,\\
 H_d(z) &=&- (z+1) H(z) H'(z)\,, \\
  H_{2 d}(z) &=& -(1+z) H(z) H_d'(z)\,, \\
  H_{3d}(z) &=& -(1+z) H(z)  H_{ 2 d}'(z)\,,\\
  H_{4d}(z) &=& -(1+z) H(z)  H_{ 3 d}'(z)\,,
\end{eqnarray}
 where $g(z)=\exp^{3 \int_0^z \frac{w(x)+1}{x+1} \, dx}$, and $w(z)$ parameterizes the dark energy EOS.
We have
\begin{eqnarray}
  \lim_{z->0}H_d(z) &=& -H_0 \left(1+q_0 \right)\label{deparcosmo1}\,, \\
    \lim_{z->0}H_{2 d}(z) &=& H_0^3 \left(j_0 + 3 q_0 + 2\right)\label{deparcosmo2}\,, \\
    \lim_{z->0}H_{3 d}(z)&=& H_0^4 \left(s_0 - 4 j_0 - 3 q_0 (q_0 + 4) - 6\right)\label{deparcosmo3}\,, \\
  \lim_{z->0} H_{4 d}(z) &=& H_0^5 \left(l_0 - 5 s_0 + 10 (q_0 + 2) j_0 +\right.\nonumber\\ &+&  \left.30 (q_0 + 2) q_0 + 24\right)\label{deparcosmo4}\,.
\end{eqnarray}
 It is worth noting that the  possibility of inverting Eqs.  (\ref{deparcosmo1}--\ref{deparcosmo4}) strongly
 depends on the number of cosmographic parameters we are working with and on how many parameters enter the
 dark energy EOS. For instance, if we expect that only two cosmographic parameters, $(q_{0}, j_{0})$, are well
 constrained, it is possible to derive information about a constant dark energy model and estimate $\Omega_m$ as
\begin{eqnarray}
\Omega_{m}(q_{0}, j_{0}) &=& \frac{2 (j_{0} - q_{0} - 2 q_{0}^{2})}{1 + 2 j_{0}
- 6 q_{0}}\, , \nonumber \\
w_{0}(q_{0}, j_{0}) &=& \frac{1 + 2 j_{0} - 6 q_{0}}{-3 + 6 q_{0}}\,.
\end{eqnarray}
If $\Omega_{m}$ is considered as a free parameter, then  with the same
cosmographic parameters, $(q_{0}, j_{0})$, we can also derive some information
about a dynamical dark energy model. In our investigation we preferred this
conservative approach. All the statistical properties of the dark energy
parameters (median, error bars, etc.) can be directly extracted from the
cosmographic samples we have obtained from the MCMC analysis. However, it is
worth noting that since the map described by Eqs.
(\ref{deparcosmo1}--\ref{deparcosmo4}) is  non-linear, the equations admit
multiple solutions for any assigned n-fold ($q_0$,$j_0$,$s_0$,...): to improve
the maximum likelihood estimate, we  incorporated the restrictions on the EOS
parameters coming from cosmography by constructing a sort of constrained
optimizer within the MCMCs.  In this analysis we considered the
Chevalier-Polarski Linder (CPL) model \cite{cpl1}, \cite{cpl2}, where

 \begin{equation}
w(z) =w_0 + w_{1} z (1 + z)^{-1} \,,
\label{cpleos}
\end{equation}
where $w_0$ and $w_1$ are free, fitting parameters, characterized by the property that
\begin{eqnarray}
&&\lim_{z \to 0}w(z)=w_0\,\\
&&\lim_{z \to \infty}w(z)=w_0+w_1\,.
\end{eqnarray}
This parametrization  describes a wide variety of scalar field  dark energy models and therefore
achieves a good compromise to construct a model independent analysis.
\begin{figure}{\includegraphics[width=6 cm, height=5cm]{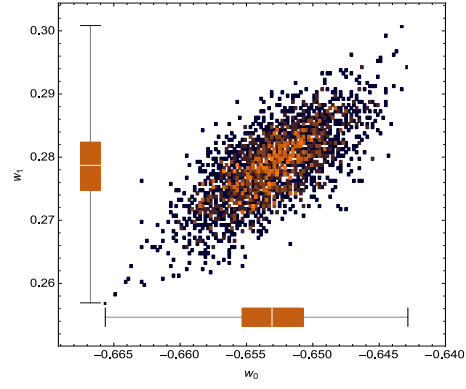}}
\caption{Confidence regions in the ($w_0$-$w_1$) plane, as provided by  Cosmography II, for the CPL parametrization of the dark energy.  The inner brown region defines the 2\,$\sigma$ confidence level. These parameters  are well constrained, and the values $w_0=-1$, and $w_1=-1$  (which characterize the standard $\Lambda$CDM model) are statistically not favorable.  The bottom and top of the diagrams plotted on the axes correspond to the 25th and 75th percentile, and the band near the middle of the box is the median.}
 \label{prova}
\end{figure}
The results of the cosmographic analysis allow us to infer the values of $w_0$ and $w_1$, thus providing
constraints on the dynamical nature of the dark energy. The EOS is evolving, as illustrated, for example, in
Fig. \ref{prova} compared to the CPL model,  thus reflecting,  at $1 \sigma$, the possibility of a deviation
from the $\Lambda$CDM cosmological model, as has been indicated by previous investigations (see, for instance, Paper 1)
\subsection{Precision cosmography from generalized Pad\'e approximation: stronger constraints for the dark energy parametrization }
Pad\'e approximation generalizes the  Taylor series expansion of a function $f(z)$: it is well known that  if
the series converges absolutely to an infinitely differentiable function, then the series defines the function
uniquely and the function uniquely defines the series. The Pad\'e approach provides an approximation for $f(z)$
through rational functions.  As an illustrative example we  consider a given power series
\begin{equation}
R(z)=\frac{p(z)}{q(z)}\,,
\end{equation}
where
\begin{eqnarray}
p(z)=\sum_{i=0}^{m} a_1 z^i\,,\\
q(z)=\sum_{i=0}^{n} b_i z^i\,,
\end{eqnarray}
where $m \leq n$.
The rational function $R^{n}_{m}(z)$ is a Pad\'e approximation to the series $f(z)$ and
\begin{eqnarray}
f(z)-R^{n}_{m}(z)= \emph{O}(z^{m+n+1}),\,
\label{pade}
\end{eqnarray}
that is, the lowest order monom in the difference \[f(z) q(z)-p(z)\] is on the order of $m+n+1$.
Equation (\ref{pade}) imposes some requirements on $R$ and its derivatives:
\begin{eqnarray}
&&R^{n}_{m}(0)=f(0)\,\\
&&\frac{d^k}{d z^k} R^{n}_{m}\|_{z=0} = \frac{d^k}{d z^k} f \|_{z=0}\,,
\label{pade2}
\end{eqnarray}
where $k=1,...m+n$.
The Eqs. (\ref{pade}, \ref{pade2}) provide $m+n+2$ equations for the unknowns $a_0,...a_m$, $b_0,...b_n$.
Since this system is, obviously, undetermined, the normalization $b_0=1$ is generally used.
 A long-standing interest in rational fractions and related topics (such as the Pad\'e approximation) is
 observed in pure mathematics, numerical analysis,  physics, and chemistry. There is a  growing interest
 today in applying the  Pad\'e approximation technique to the accelerated Universe cosmology to
 investigate the nature of dark energy \cite{nesseris}, \cite{gruber}, \cite{aviles}, and \cite{liu}.
 To construct an accurate
 generalized   Pad\'e approximation  of the distance modulus and investigate the implications on the
 cosmography, we here started from a two-parameter $a,b$  Pad\'e approximation  for the deceleration parameter
 $q(z)$, and obtained $H(z)$ and the luminosity distance $d_L(z)$ according to the relations,
 here we assume a flat cosmological model,
 \begin{eqnarray}
&& H(z)=H_0 \exp (\int^{z}_0 \frac{1+ q(u)}{1+u} du)\\
&& d_L(z)= \left(1+z\right) \int_{0}^{z} \frac{du}{H(u)}\,.
\label{Hdlq}
 \end{eqnarray}
  For $H(z)$ we obtain a power-law approximation with fixed exponent, and for $d_L$ an exact analytical
  expression in terms of  the Gauss hypergeometric function, $_2F_1$:
  \begin{eqnarray}
&& d_L(z)=\frac{c}{100 h} \frac{(z+1) (\beta +1)^{\frac{\alpha +\beta  \gamma }{\beta  \delta }}}{\gamma }\times\\
  && \left[(z+1)^{\gamma } \, _2F_1\left(\frac{\gamma }{\delta },\frac{\alpha +\beta
   \gamma }{\beta  \delta };\frac{\gamma }{\delta }+1;-(z+1)^{\delta } \beta \right)-\right.\nonumber\\
&&\left.   _2F_1\left(\frac{\gamma }{\delta },\frac{\alpha +\beta  \gamma }{\beta  \delta
   };\frac{\gamma }{\delta }+1;-\beta \right)\right]\nonumber\,,
\label{dlpade}
 \end{eqnarray}

 \begin{figure}{ \includegraphics[width=8 cm, height=5 cm]{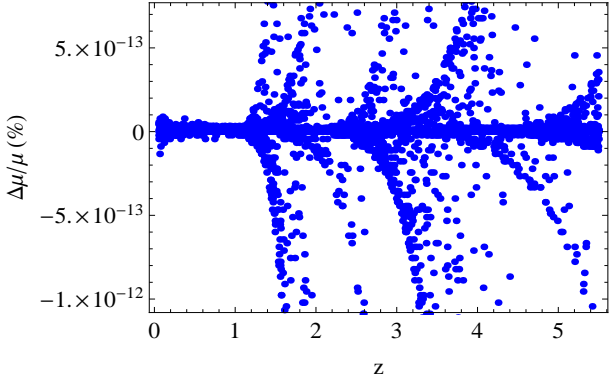}}
        \caption{ Percentage error between a randomly generated high-redshift  $\Lambda$CDM  Hubble diagram and its
        generalized   Pad{\'e} approximation  : this approximation is more accurate. }
        \label{padenesseris}
\end{figure}
where $\alpha$, $\beta$, $\gamma$ and $\delta$ are fitting parameters. This {extended } Pad\'e approximation
works even better than the the {original} approximation, as shown in Fig. (\ref{padenesseris}), where we
 evaluate the relative error between a randomly generated high-redshift  $\Lambda$CDM Hubble diagram and its extended
 Pad\'e approximation.
The parameters  $\alpha$, $\beta$, $\gamma,$ and $\delta$ have been constrained using the same statistical
analysis described previously, that is, by implementing the MCMC  simulations and simultaneously computing the
full probability density functions of  these parameters. It is, moreover,  possible to map the extended
Pad\'e parameters into the cosmographic parameters, allowing a refined statistics of its specific parameters
(especially $q_0$ and $j_0$) and, at last, stronger constraints on  the dark energy EOS. The values $q_0>0$ are
significantly ruled out, and the indications favoring  the value $j_0\not= 1$ are much stronger than in the
cosmographic analysis. Equations (\ref{deparcosmo1}---\ref{deparcosmo4}) allow projecting the constraints obtained
from the Pad\'e analysis on the space of the EOS parameters: a dynamical dark energy is strongly supported by this
analysis, as shown in Table (\ref{tabpadecpl}), compared to the CPL parametrization.
\begin{table}
\begin{center}
\begin{tabular}{|c|c|c|}
 \hline
\hline
  Parameter &$w_0$&$w_1$\\
  \hline
  \hline
  Best fit & $-0.6150$&$ 0.329$\\
  Mean & $-0.6149$&$0.32930$ \\
  2 $\sigma$ & $(-0.6180, -0.6118)$&$(0.325, 0.333)$\\
  \hline
  \hline
\end{tabular}
\end{center}
\caption{Median (best fit), mean values, and
confidence regions in the ($w_0$-$w_1$) plane, by  projecting the constraints from the
generalized  Pad\'e approximation on the space of the EOS parameters: here we consider as
an illustrative example the case of the CPL parametrization.}
\label{tabpadecpl}
\end{table}

\section{Conclusions}
We investigated the dynamics of the Universe by using a cosmographic approach: we
performed a high-redshift analysis that allowed us to set constraints on the
cosmographic expansion up to the fifth order, based on the Union2 SNIa data set,
the GRB Hubble diagram, constructed  by calibrating the correlation between the
peak photon energy, $E_{\mathrm{p, i}}$, and the isotropic equivalent radiated
energy, $ E_{\mathrm{iso}}$, and Gaussian priors on the distance from the BAO,
and the Hubble constant $h$ (these priors were included to help break the
degeneracies among model parameters). Our statistical analysis was based on MCMC
simulations to simultaneously compute the regions of confidence of all the
parameters of interest. Since methods like the MCMC are based on an algorithm
that randomly probes the parameter space, to improve the convergence, we  imposed
some constraints on the series expansions of $H(z)$ and $d_{L}(z)$, requiring
that in each step of our calculations $d_{L}(z) > 0$ \,, and $H(z) > 0$. We
performed the same MCMC calculations, first considering the SNIa Hubble diagram and the BAO
data sets or the GRBs Hubble diagram, and the BAO  data sets separately (Cosmography Ia and
Ib, respectively), and then constructing an overall data set joining them
together (Cosmography II). Our MCMC method allowed us to obtain constraints on
the parameter estimation, in particular for higher order cosmographic parameters
(the jerk and the snap). The deceleration parameter confirms the current
acceleration phase; the estimation of the jerk reflects at $1 \sigma$  the
possibility of a deviation from the $\Lambda$CDM cosmological model. Moreover, we
investigated implications of our results for the reconstruction of the dark
energy EOS by comparing  the standard  technique of cosmography with an
alternative approach based on  generalized Pad\'e approximations of the same
observables. Owing to the better convergence properties of these expansions, it
is possible to improve the constraints on the cosmographic parameters and also on
the dark matter EOS: our analysis indicates that at the $1 \sigma$ level the dark
energy EOS is evolving.

\subsection*{Acknowledgments}
MD is grateful to the INFN for financial support through the Fondi FAI GrIV.
EP acknowledges the support of INFN Sez. di Napoli  (Iniziative Specifica QGSKY and TEONGRAV).
LA  acknowledges support by the Italian Ministry for Education,
University and Research through PRIN MIUR 2009 project on "Gamma ray
bursts: from progenitors to the physics of the prompt emission process." (Prot. 2009 ERC3HT).

\bibliographystyle{aa}

\begin{thebibliography}{99}
\expandafter\ifx\csname natexlab\endcsname\relax\def\natexlab#1{#1}\fi
\bibitem[{Akarsu et al.}\, {2015}]{akarsu15}
Akarsu, \"O, Bouhmadi-L\'opez, M., Brilenkov, M., Brilenkov, R., Eingorn, M.,  Zhuk, A., 2015, JCAP, 7, 38
\bibitem[{Amati et al.}\, {2008}]{Amati08}
Amati, L., Guidorzi, C., Frontera, F., et al., 2008, MNRAS, 391, 577
\bibitem[{Aubourg et al. (BOSS Collaboration)}\, {2015}]{Aubourg14}
Aubourg., E., et al. (BOSS Collaboration), 2015, Phys. Rev. D 92, 123516
\bibitem[{Aviles et al.}\,{2014}]{aviles}
Aviles, A., Bravetti, A., Capozziello, S., Luongo, O., 2014, Phys. Rev. D, 90, 043531
\bibitem[{Bouhmadi-Lopez et al.}\, {2010}]{salv}
Bouhmadi-Lopez, M., Capozziello, S., Cardone, V. F., 2010, Phys.Rev D,  82, 103526
\bibitem[{Astier et al.}\, {2006}]{SNLS}
Astier, P., Guy, J., Regnault, N., Pain, R., Aubourg, E., et al.
2006, A\&A, 447, 31
\bibitem[{Bond et al.}\, { 1997}]{B97}
Bond, J.R., Efstathiou, G., Tegmark, M., 1997, MNRAS, 291, L33
\bibitem[{Busca et al.}\, {2013}]{busca}
Busca, N. G., Delubac, T., Rich, J., Bailey, S., Font-Ribera, A., et al. 2013, A\&A, 552,18
\bibitem[{Capozziello, Lazkoz, and Salzano}\, {2011}]{salzcosmo}
Capozziello, S., Lazkoz, R., Salzano, V., 2011, Phys. Rev. D, 84,  124061
\bibitem[{Chen and Ratra}\, {2011}]{chen}
Chen, G., Ratra, B., 2011,PASP, 123,1127
\bibitem[{Chevallier and Polarski}\, {2001}]{cpl1}
Chevallier, M., Polarski, D., 2001, Int. J. Mod. Phys. D, 10,  213-224
\bibitem[{Demianski, Piedipalumbo and Rubano}\, { 2011}]{MEC10}
Demianski, M., Piedipalumbo, E., Rubano, C., 2011, MNRAS, 411, 1213
\bibitem[{Demianski and Piedipalumbo}\, { 2011}]{ME}
Demianski, M., Piedipalumbo, 2011, MNRAS,  415, 3580
\bibitem[{Demianski et al.}\, {2012}]{MEcosmo}
Demianski, M., Piedipalumbo, E., Rubano, C. and Scudellaro, P. , 2012, MNRAS, 426, 13961415.
\bibitem[{Efstathiou and Bond}\, {1999}]{EB99}
Efstathiou, G., Bond, J.R., 1999, MNRAS, 304, 75
\bibitem[ {Eisenstein and Hu}\, {1998}]{EH98}
Eisenstein,D.J.,  Hu, W., 1998, ApJ, 496, 605
\bibitem[{Farroq, Mania and  Ratra}\, {2013}]{farooqa}
Farroq, O., Mania, D., Ratra, B., 2013, ApJ, 764,13
\bibitem[{Farroq and Ratra}\, {2013}]{farooqb}
Farroq, O., Ratra, B., 2013, ApJ, 766, L7
\bibitem[{Liu and Wei}\, {2015}]{liu}
Liu, J., Wei, H., 2015, Gen. Rel. Grav.47, 141
\bibitem[{Gao et al.}\, {2010}]{gao}
Gao, H., Liang, N., Zhu, Z.-H., 2010, arXiv:1003.5755
\bibitem[ {Gruber and Luongo }\, {2014}]{gruber}
Gruber, C., Luongo, O., 2014, Phys. Rev. D, 89, 103506
\bibitem[{Hinshaw et al.}\, {2013}]{hinshaw}
Hinshaw, G., Larson, D., Komatsu, E., Spergel, D. N., Bennett, C. L., 2013, ApJ S, 19,208
\bibitem[{Lazkoz et al.}\,{2010}]{escamillarivera}
Lazkoz, R., Lazkoz, Alcaniz, J., Escamilla-Rivera, C.,  Salzano, V., Sendra, I., 2010, JCAP, 12,5
\bibitem[ {Li et al. }\, {2008}]{Li08}
Li, H., Su, M., Fan, Z., Dai, Z., Zhang, X., 2008, Phys. Lett. B,
658, 95
\bibitem[{Liang et al. }\,{2008}]{Liang08}
Liang, N., Xiao, W. K., Liu, Y., Zhang, S. N., 2008, ApJ, 685, 354
\bibitem[ {Liu and Wei }\,{2014}]{Liu14}
Liu, J., Wei, H., 2014, arXiv1410.3960L
\bibitem[{Linder }{2003}]{cpl2}
Linder, E.V., 2003, Phys. Rev. Lett., 90, 091301
\bibitem[{Moresco et al.}\, {2016}]{moresco}
Moresco, M., Pozzetti, L.; Cimatti, A.; et al. 2016,
A $6\%$ measurement of the Hubble parameter at $z\sim0.45$: direct evidence of the epoch of
cosmic re-acceleration, eprint arXiv:1601.01701
\bibitem[{Nesseris and Garcia-Bellido }\,{2013}]{nesseris}
Nesseris, S., Garcia-Bellido, J., 2013, Phys.Rev. D, 88, 063521
\bibitem[ {Peebles and Ratra }\,{1988a}]{Peebles88a}
Peebles, P. J. E., Ratra, B., ApJ, 1988, 325,17
\bibitem[ {Ratra and Peebles}\,{1988b}]{Peebles88b}
 Ratra, B., Peebles, P. J. E.,  Phys. Rev. D, 1988, 37, 3406
\bibitem[ {Planck Collaboration }\,{2013}]{PlanckXXVI}
Planck Collaboration, {\em Planck 2013 results. XXVI. Cosmological parameters},
 arXiv:1303.5076
\bibitem[{Planck Collaboration }\,{2015}]{PlanckXIII}
Planck Collaboration, {\em Planck 2015 results. XIII. Cosmological parameters},
arXiv:1502.01589
\bibitem[{Percival et al. }\, {2010}]{P10}
Percival, W.J., Reid, B.A., Eisenstein, D.J., Bahcall, N.A., Budavari, T., et al., 2010, MNRAS,  401, 2148
\bibitem[{Perlmutter et al. }\, {1999}]{per+al99}
Perlmutter, S., Aldering, G., Goldhaber, G., Knop, R. A., Nugent, P., et al. 1999, ApJ, 517, 565
\bibitem[ {Riess et al. }\, {1998}]{Riess}
Riess, A.G., Filippenko, A.V., Challis, P., Clocchiatti, A., Diercks, A., et al., 1998, ApJ, 116,1009
\bibitem[ {Riess et al.}\, {2007}]{Riess07}
Riess, A.G., Strolger, L.G., Casertano, S., Ferguson, H.C.,
Mobasher, B., et al., 2007, ApJ, 659, 98
\bibitem[ {Riess et al.}\, {2009}]{shoes}
Riess, A.G., Macri, L., Li, W., Lampeitl, H., Casertano, S., et al. 2009, ApJ, 699, 539
\bibitem[ {Riess et al. }\,{2011}]{Riess11}
Riess, A. G., Macri, L., Casertano, S.,  Lampeitl, H., Ferguson, H., C., 2011, ApJ,  730, 119
\bibitem[ {Sahni et al. }\, {2003}]{SF}
Sahni, V., Saini, T.D., Starobinsky, A.A., Alam, U., 2003, JETP Lett., 77, 201;
U. Alam, V. Sahni, T.D. Saini, A.A. Starobinsky, 2003, MNRAS, 344, 1057
\bibitem[ {Samushia and Ratra }\,{2010}]{Samushia}
 Samushia, L., Ratra, B., 2010,  ApJ,  714, 1347
 \bibitem[{Sievers et al.}\, {2013}]{sievers}
 Sievers, J. L., Hlozek, R. A.; Nolta, M. R., Acquaviva, V., Addison, G. E., 2013, JCAP, 60,1310
\bibitem[ {Suzuki et al.  }\,{2012}]{Union2.1}
 Suzuki et al. (The Supernova Cosmology Project),  2012,  ApJ,  746, 85
 \bibitem[ {Tsutsui et al. }\, {2009}]{Ts09}
Tsutsui, R., Nakamura, T., Yonetoku, D., Murakami, T., Tanabe, S., et
al., 2009, MNRAS, 394, L31-L35
\bibitem[ {Visser }\, {2004}]{Visser}
Visser, M., 2004, Class. Quant. Grav., 21, 2603
\bibitem[ {Vitagliano et al. }\, {2010}]{vitagliano}
Vitagliano, V., Xia, J.Q., Liberati, S., Viel, M., 2010, JCAP, 3, 005
\bibitem[ {Wang }\, {2008}]{Wang09}
Wang, Y., 2008, Phys. Rev. D, 78, 123532
\bibitem[ {Wang, J.S., et al. }\,{2015}]{Wang15}
 Wang, J., Deng, J.S., and  Qiu, Y.J., 2008,  Chin. J. Astron. Astrophys., 8, 255
 \bibitem[{Wang, F.Y. et al.}\,{2015}]{wangrev}
 Wang, F. Y., Dai, Z. G.;,Liang, E. W., 2015, NewAR, 67,1
  \bibitem[{Wang, F.Y. et al.}\, {2016}]{wangjs16}
Wang, J. S., Wang, F. Y., Cheng, K. S., Dai, Z. G., 2016, A\&A, 585, 68
\bibitem[ {Wei }\, {2010}]{wei}
Wei, H., 2010, JCAP,  8, 20
\bibitem[ {Wood\,-\,Vasey et al. }{2007}]{ESSENCE}
Wood\,-\,Vasey, W.M., Miknaitis, G., Stubbs, C.W., Jha, S., Riess,
A.G., et al., 2007, ApJ, 666, 694
\bibitem[{Zhan et al.}\, {2016}]{zhang16}
Zhang, M.-J., Li, H., Xia, J.-Q., 2016, Cosmographic analysis from distance indicator and dynamical redshift drift,
eprint arXiv:1601.01758
\end{thebibliography}

\end{document}